\documentclass[referee]{aa}
\usepackage{psfig}
\onecolumn

\begin{document}
\thesaurus{     11.04.1   % Galaxies: distances and redshifts
                11.04.2   % Galaxies: dwarf
                11.06.2   % Galaxies: fundamental parameters
                11.09.5   % Galaxies: irregular
                11.12.1   % Local Group
                11.19.5   % Galaxies: stellar content
          }

\title{The Sagittarius Dwarf Irregular Galaxy (SagDIG): distance and
star formation history\footnote{Based on observations made with the 2.5
m Nordic Optical Telescope operated on the island of La Palma by NOT
S.A. in the Spanish Observatorio del Roque de Los Muchachos of the
Instituto de Astrof\'\i sica de Canarias.}}

\titlerunning{Distance to Sagittarius Dwarf Irregular Galaxy}

\author{I. Karachentsev\inst{1} \and A. Aparicio\inst{2} \and L. Makarova\inst{1}}

\authorrunning{I. Karachentsev et al.}

\institute{Special Astrophysical Observatory, Russian Academy of Sciences,
      N.Arkhyz, KChR, 357147, Russia
\and Instituto de Astrof\'\i sica de Canarias, E-38200 La Laguna,
      Canary Islands, Spain}

\date{}

\maketitle

\begin{abstract}

The distance, star formation history and global properties of the Local Group
dIrr galaxy SagDIG are derived based on an [$I-(V-I)$] colour--magnitude
diagram of $\sim1550$ stars.  A distance of $1.06\pm 0.10$ Mpc is obtained
from the $I$ magnitude of the TRGB. This corresponds to 1.17 Mpc to the
barycenter of the Local Group and 1.34 to M 31, being DDO 210, at 0.35 Mpc,
the nearest galaxy to SagDIG. The metallicity is estimated
from the colour of the RGB to be [Fe/H] $=-2.45\pm 0.25$. SagDIG is hence a
probable member of the Local Group and a candidate for the lowest-metallicity
star forming galaxy known.

The radial density profile of the galaxy has been obtained together
with other integrated properties (magnitude, colour and central surface
density). The galaxy density profile is fitted by an exponential law of
scale length $27\farcs1$, corresponding to 140 pc.

The star formation history of SagDIG has been analysed, based on
synthetic colour--magnitude diagrams. The galaxy is currently in a high
star formation activity epoch, forming stars at a rate about 10 times
greater than the average for its entire life. This is a common feature of
galaxies classified as dIrrs.


\end{abstract}

\section{Introduction}

The Sagittarius dwarf irregular galaxy (SagDIG) is a blue, low surface
brightness galaxy which was found on the ESO and SERC survey plates by
Cesarsky et al. (1977) and Longmore et al. (1978). Both teams detected
the galaxy in the 21-cm line with a negative radial velocity which
indicates its probable membership in the Local Group. Its brightest blue
stars are asymmetrically distributed, being concentrated on the eastern
side of the galaxy. Based on their apparent magnitude, $B \sim 18.5$,
Cesarsky et al. (1977) estimated the galaxy distance modulus to be
$25\pm 1$. Skillman, Terlevich, \& Melnick (1989) estimated the metallicty of
its ionized gas to be 3\% of the solar value and Strobel, Hodge, \& Kennicutt (1991) give an $H_\alpha$ map of
the galaxy. Later on, Lo, Sargent \& Young (1993) and Young \& Lo
(1997) undertook detailed investigations of SagDIG in the H~{\sc i} line with
the VLA and showed that its velocity field is dominated by chaotic
motions rather than by rotation ($V_{\rm rot}<2$ km/s). For the H~{\sc i} mass
and the total (virial) mass of SagDIG Lo et al. (1993) derived $M_{\rm
HI}= 7.4\cdot10^6\ M_\odot$ and $M_{\rm vt}= 3.8\cdot10^7\ M_\odot$
respectively.

Surprisingly, SagDIG remains the least optically studied dIrr galaxy
among the Local Group members (see, for example, bibliography in Gallart
et al. 1996a and Mateo 1998). Even the existing estimates of its
integrated magnitude are spread over a range of 15.5 -- 13.8. The
scarcity of studies devoted to SagDIG encouraged us to undertake the
detailed CCD imaging of the galaxy that we present here. The paper is
organized as follows. Section 2 gives a short description of the
observations. In Sec. 3 the colour--magnitude diagram (CMD) is described
and the distance and metallicity of the galaxy are derived from the tip
of the red giant branch (TRGB) and the colour of the red giant branch
(RGB), respectively. In Sec. 4 the radial distribution of stars is
studied and the integrated magnitude obtained. The star formation
history is derived in Sec. 5. Section 6 presents a summary of the global
properties of SagDIG. Finally, the conclusions of the paper are
summarized in Sec. 7.

\section{Observations and data reduction}

SagDIG was observed on 1997 July 28 with the HiRAC CCD camera of the
2.56 m NOT at Roque de los Muchachos Observatory in La Palma island
(Canary Islands, Spain). The seeing was $0\farcs8$. A 2048$\times$2048
CCD detector was used binned to $2\times 2$ and provided a total field
of 3.7 $(')^2$ with a resolution of $0.22('')$/pixel after rebinning. Two
frames were obtained in both the $V$ and $I$ bands with total exposures 2000
s in $V$ and 1800 s in $I$. In order to estimate foreground galactic
star number counts another field (8$'$ to the east) was imaged with
exposures of 1200 s ($V$) and 1000 s ($I$). Standard stars from Landolt
(1992) were used for calibration. Details of the calibration can be
found in Aparicio, Tikhonov \& Karachentsev (1999). A $V$ band frame of
SagDIG is shown in Figure 1.

The derived images were processed with MIDAS implementation of the
DAOPHOT II program (Stetson 1987; Stetson, Davis \& Crabtree 1990). A
total of about 1550 stars were measured in the SagDIG frames down to a
limiting magnitude $I_{\rm lim}\sim 23.5$. Figure 2 plots the PSF fitting
errors as provided by ALLSTAR for $V$ and $I$ magnitudes.

Completeness has been analyzed using the usual procedure of artifitial star
trials (Stetson 1987). A total of 2000 artifitial stars were added to the $V$
and $I$ frames of SagDIG in several steps of 100 stars each with magnitudes
and colors in range $17\leq V\leq 25$ and $-0.4\leq(V-I)\leq 3.2$. Stars were
considered as recovered if they were found both in $V$ and $I$ with
magnitudes not exceding 0.75 mag brighter than the initial, injected
ones. The completeness curves for blue and red stars are shown in Figure 3.

Besides stellar photometry we
also carried out aperture photometry of the galaxy in circular
diaphragms that allows  the total integrated magnitude and
colour to be measured.

\section{Distance and metallicity from the colour--magnitude diagram}

The $I$ vs. $(V-I)$ CMD of the resolved stars in the SagDIG frame is shown in
Figure 4. SagDIG lying close to the direction of the Galactic center, has a
CMD heavily contaminated by foreground stars. Figure 5 shows the CMD of the
aforementioned nearby field. In both figures, only stars with $SHARP$ and
$CHI$ $ALLFRAME$ parameters in the intervals $-1.5\leq SHARP\leq 1.5$ and
$CHI\leq 2$ in both filters have been plotted, being 1513 in Fig. 4 and 648
in Fig. 5. Comparison of both CMDs shows that for $(V-I)<0.6$, the former is
almost free from foreground stars, while most of the stars redder than that
value and brighter than $I\sim 21$ must be Galactic members. Two features can
be recognized to originate in SagDIG: the blue sequence with $(V-I)<0.4$,
extending up to $I\sim 19.5$ and the sequence of red stars at $(V-I)\sim 1.0$
to 1.4, extending up to $I\sim 21.5$. The former is identified as being
produced by young main sequence and He-burning blue-loop stars while the
latter can be recognized to be the red giant branch (RGB), and should also
contain asymptotic giant branch (AGB) stars.

\subsection{Distance}

The distance to SagDIG can be obtained from the tip of the RGB (TRGB),
which for metal poor systems like SagDIG (see below) can be assumed to
be at $M_{\rm I}=-4.0$ (Da Costa \& Armandroff 1990).  

The magnitude of the TRGB has been obtained applying a Sobel filter (kernel
[--1,0,+1]; see Myler, \& Weekes 1993) to the luminosity function of stars
with $1.0<(V-I)<1.7$. To minimize the effects of foreground contamination,
only the stars at less than 1$'$ from the center of SagDIG have been used
(see below for the dimensions of the galaxy). The resulting luminosity
function and Sobel filtered luminosity function are shown in Figure 6. The
TRGB corresponds to the peak at $I=21.38$ (bottom panel), the one at at
$I=22.0$ being produced by a density fluctuation inside the RGB. The error
can be estimated as 1/2 of the peak width at 62\% of its maximum and turns
out to be $\pm 0.15$.

We have adopted a Galactic reddening for SagDIG of $E(B-V)=0.12$ or
$A_I=1.8\times E(B-V)=0.22$ from the IRAS/DIRBE map (Schlegel, Finkbeiner \&
Davis 1998). The above values yield a distance modulus of $(m-M)_0=25.13\pm
0.20$ or $D_{\rm MW}=1.06\pm 0.10$ Mpc, where the standard error includes
uncertainties in the exact location of the TRGB, extinction, and photometric
calibration. SagDIG is at $D_{\rm LG}=1.17$ Mpc from the barycenter of the
Local Group and at 1.34 from M31. The closest galaxies to SagDIG are DDO 210,
at 0.35 Mpc and NGC 6822, at 0.56 Mpc. Hence SagDIG seems to be a rather
isolated galaxy in the periphery of the Local Group.

The average magnitude of the three brightest stars in a galaxy is a simple,
frequently used method to estimate the distance. However, for low surface
brightness galaxies without very young stars this method usually leads
essentially to overestimating the distance. For ilustrative purposes, we have
obtained the distance to SagDIG by this method also. The mean apparent
magnitude of the three brightest blue stars can be estimated from our $VI$
photometry using the relation $(B-V)=0.83\times (V-I)$ obtained from blue
[$(V-I)\leq 0.6$] standard stars of Landolt (1992). Applying this to the
stars with $(V-I)\leq 0.6$ in Fig. 4 the mean apparent magnitude of the three
brightest ones is $\langle B(3B)\rangle \simeq 19.69$, or $\langle
B(3B)\rangle \simeq 19.82$ if the brightest blue star [$I=18.82$,
$(V-I)=0.46$] is assumed to be a foreground contaminator and
neglected. Moreover, the integrated $B$ magnitude of the galaxy can be
estimated using the values $V_{\rm T}=13.77$ and $(V-I)_{\rm T}=0.65$ derived
in Sec. 4 and the relation $(B-V)_{\rm T}=0.85\times (V-I)_{\rm T}-0.20$ by
Makarova, \& Karachentsev (1998). It results $B_{\rm T}\simeq 14.12$.  Using
the standard relation $(m-M){_0}=1.51\times \langle B(3B)\rangle -0.51\times
B_{\rm T}-A_{\rm B}+4.14$ from (Karachentsev \& Tikhonov 1994) with
$A_{B}=0.51$ it is obtained $(m-M)_0=26.16$ (26.36) if the brightest star is
(is not) considered, which is about 1 mag more than the distance modulus via
TRGB or more than 50\% larger in the distance. It must be noted that Cesarsky
{\it et al.} (1977) obtained $\langle B(3B)\rangle \simeq 18.5$, which would
result in $(m-M)_0=24.36$ and that these authors claimed a distance modulus
of $25\pm1$, close to the TRGB estimate. However Cesarsky {\it et al.}
estimate was based on rough eye photometry on photographic plates and could
well be affected of severe blending.

\subsection{Metallicity}

The metallicity of SagDIG was measured by Skillman, Terlevich, \& Melnick
(1989) from a low surface brightness HII region in the galaxy. They obtained
$Z=0.0006$. Alternatively, we have estimated the mean metallicity of the
stars in SagDIG from the mean $(V-I)$ colour of the RGB at $M_I= -3.5$,
i.e. 0.5 mag fainter than the TRGB.  At this level we have a mean
$(V-I)_{0,-3.5} = 1.17\pm 0.05$ assuming a reddening of $E(V-I)=0.16$. Lee et
al. (1993) provided a calibration for the metallicity of the RGB based on the
$(V-I)$ colours at $M_I=-3.5$ for Galactic globular clusters (Da Costa, \&
Armandroff, 1990): [Fe/H] $= -12.64+12.61(V-I)_{0,-3.5} -3.33(V-I)^2
_{0,-3.5}$. Using this calibration we obtain a value for the mean metallicity
of [Fe/H] $=-2.45\pm0.25$. This value is quite smaller than the Skillman
et al.'s one. However, it must be noted that it corresponds to intermediate
and old stars in the galaxy and, more important, that Lee {\it et al.}'s
relation is used in extrapolation, since the least metallic globular cluster
(M 15) used by Da Costa, \& Armandroff (1990) has [Fe/H]$=-2.17$. In any
case, SagDIG lies at the extreme metal-poor end of dwarf galaxies. Figure 7
shows the RGB fiducials of Da Costa, \& Armandroff (1990) overplotted to the
distance and reddening corrected CMD of SagDIG.

\section{Radial density profile and integrated photometry}

The radial density profile of SagDIG has been obtained by counting
stars in circular concentric annuli from the apparent optical  center
of the galaxy. Figure 8 shows the results. It is apparent from this
figure that the galaxy vanishes at $r\simeq 120 ''$, the background
level corresponding to about 0.016 stars arcsec$^{-2}$. From the nearby
companion field that we have taken, a similar but slightly lower value
of 0.013 stars arcsec$^{-2}$ is obtained.

Subtracting the former value of the background to the density profile, an
exponential law has been fitted to the region $20''\leq r <100''$,
which results in a scale length of $27\farcs1$ or 140 pc.

Total integrated magnitudes and colours have also been obtained. To reduce
the effects of foreground contamination, we have first remove all the stars
with $V\leq 20$ and $0.6\leq (V-I)\leq 1.2$. The sky level has been
approximated by a 2-dimension polynomial, using regions with few stars near
the edges of the images. Then, integrated photometry of SagDIG has been
performed with increasing circular apertures. The galaxy magnitude in each
band has then been measured as the asymptotic value of the derived growth
curve. The $(V-I)$ colour has been determined as the difference of total
magnitudes in each band. The results are $V_T=13.77$ and $(V-I)_T=0.65$.

Figure 9 shows the CMDs for 2 circular regions centered in the galaxy center
and radii 60$''$ and 120$''$, 
respectively. It visualizes the relative contribution of SagDIG and
foreground stars for increasing radial distances.
 
\section{The star formation history of SagDIG}

The heavy foreground contamination of SagDIG makes the analysis of the
SFH difficult. It is based on the distribution of stars in the CMD and
uses star counts in low populated areas of the CMD also (see Gallart et
al. 1999). Fortunately enough, the bluest part of the SagDIG CMD,
populated by youngest stars, is free from contamination and can provide
information about the very recent SFH. Besides this, the RGB area can be
used, after appropriate correction of foreground contamination to
estimate the averaged SFR for ages older than $\sim 1$ Gyr (Aparicio,
Bertelli  \& Chiosi 1999).

To start with, a qualitative idea of the stellar ages can be obtained from a
glance at Figure 10, which shows the CMD of SagDIG corrected from the
reddening and distance given in Section 3. Five isochrones from the Padua
library (see Bertelli et al. 1994 for the key reference) have been
over-plotted with metallicity $Z=0.0004$ and ages 20, 50, 200 Myr and 1 and
10 Gyr. The RGB and AGB are shown for the latter. Only the stars within a
radius of $2'$ from the center of the galaxy have been plotted, which
approximately corresponds to the maximum extension of the galaxy (see
Sec. 4). It is worth noting that at least some of the stars in the strip
extending from $[I,(V-I)]=[-4.0,1.0]$ to $[I,(V-I)]=[-7.0,1.5]$ are probably
extended intermediate-mass AGBs belonging to SagDIG (see below). Note also
the separation between the MS and the blue-loop sequence, and that the
brightest blue stars are blue-loopers.

A simplified version of the method proposed by Aparicio, Gallart \&
Bertelli (1997b) has been employed to obtain the SFH of SagDIG. In
practice, a synthetic CMD with arbitrary, constant SFR of value
$\psi_p$, the IMF of Kroupa, Tout \& Gilmore (1993) with lower and upper
cut-offs 0.1 M$_\odot$ and 30 M$_\odot$, respectively, and a metallicity, $Z$,
taking random values from $Z_1=0.0004$ to $Z_2=0.0005$, independently
of age have been used. To avoid small number statistics effects, the
synthetic CMD have been computed with 50000 stars with $M_I\leq -2$. This
guarantees that the relevant regions of the CMD (see below) are well
populated and that the synthetic CMD does not introduce further statistical
errors to the SFH result. In Sec. 3 the metallicity of SagDIG was 
estimated at  [Fe/H] $=-2.45$. This is an extremely low value, probably
corresponding to $Z<0.0001$. However, the Padua library is not
complete for such low metallicities. For this reason the former
metallicity range has been used as representative of a very low
metallicity galaxy. Finally, since our results for the SFR can only be
an estimate, we have neglected the effects of binary stars.

The resulting synthetic CMD has then been divided into three age
intervals: $0-0.05$ Gyr, $0.05-0.2$ Gyr and $0.2-15$ Gyr. Following the
nomenclature introduced in Aparicio et al. (1997b), each of
the synthetic diagrams corresponding to the three previously defined  age
intervals will be called {\it partial model}
CMDs and any linear combination of them will be denoted as
{\it global model}. Three regions have been defined in the
observed and  partial model CMDs as shown in Fig. 10, with
the criterion that they sample different age intervals and
stellar evolutionary phases, namely the youngest blue-loops
[$(V-I)_0\leq 0.4$; $-6\leq M_{\rm I_0}<-4.5$], the MS plus young blue-loops
[$(V-I)_0\leq 0.4$; $-4.5\leq M_{\rm I_0}<-3$], and the RGB+AGB region below
intervals: the TRGB 
[$0.6\leq (V-I)_0\leq 1.5$; $-4\leq M_{\rm I_0}<-3$].  In practice, the two
youngest time intervals are sampled by blue stars only, the oldest age
included in box 2 of Fig. 10 being in fact about 0.2 Gyr. Although stars of
any age above 0.2 Gyr populate box 3 (Fig. 10), this box is in practice
dominated by low-mass stars, which are therefore older than about 1 Gyr, so
that the sampling of stars in the age interval 0.2--1 Gyr remains poor. This
interval should be solved using AGB stars brighter than the TRGB. But the low
star counts usually found in the upper AGB together with the high foreground
contamination prevent us from making any estimate based on that region. In
summary, we will give the average SFR for the 0.2--15 Gyr interval, but the
bad sampling of the 0.2--1 Gyr interval must be borne in mind.

We denote by $N_j^o$ the number of stars of the observed CMD lying in region
$j$ and by $N_{ji}^m$ the number of stars of partial model (age interval) $i$
populating region $j$. After completeness and foreground correction, $N_j^o$
take the following values: $N_1^o=16.3$, $N_2^o=82.6$, and $N_3^o=258.3$
(inner $2'$). The number of stars populating a given region in a global model
is then given by

\begin{equation}
N_j^m=k\sum_i\alpha_iN_{ji}^m
\end{equation}
\noindent and the corresponding SFR 
\begin{equation}
\psi(t)=k\sum_i\alpha_i\psi_p\Delta_i(t),
\end{equation}
\noindent where $\alpha_i$ are the linear combination coefficients; $k$ is a
scaling constant which transforms from the arbitrary units used in the
computation to final, physical units; $\Delta_i(t)=1$ if $t$ is inside the
interval corresponding to partial model $i$ and $\Delta_i(t)=0$ otherwise. In
the simple approach we are using, the $\alpha_i$ coefficients can be
analytically solved to produce $N_j^m=N_j^o$. This results in the SFRs for
the three considered intervals of time given in Table 1. The first three
lines give the SFR for the time intervals $0.2-15$ Gyr; $0.05-0.2$ Gyr and
$0-0.05$ Gyr. The three next lines give the same normalized to the area of
SagDIG, considered to extend to the $\mu_{\rm B}=25.6$ Holmberg
radius. Quoted errors have ben calculated assuming Poisson statistics in the
$N_j^o$ star counts.

Figure 11 shows the synthetic CMD corresponding to the former solution of the
SFH. No simulation of observational effects has been done. As a result, the
lowest part of the diagram is much more clearly defined than that of
Fig. 4. Note in particular the separation between MS and blue-loop stars,
which is only marginally visible in the observational CMD. Interestingly a
large amount of bright AGB stars populate the synthetic CMD, indicating, as
previously stated, that, at least, some of the stars in the strip extending
from $[I,(V-I)]=[-4.0,1.0]$ to $[I,(V-I)]=[-7.0,1.5]$ of Fig. 4 are
AGBs. Unfortunately, the strong foreground contamination prevents using these
stars to improve the SFH result.

The current SFR of SagDIG can also be estimated from the $H_\alpha$ flux
given by Strobel et al. (1991). Using our estimate of the distance and
following the procedure shown in Aparicio et al. (1999) with an upper mass
for stars $m_u=25$M$_\odot$, the current SFR results $\psi(0)=1.5\times
10^{-4}$M$_\odot$yr$^{-1}$; i.e., an order of magnitude smaller than the
value obtained from the CMD and given in Table 1 for the last 50 Myr
($\bar\psi_{0-0.05}$). The disagreement can be solved if an upper cut-off for
stellar masses of $m_u\simeq 12$ is imposed.

Summaryzing, SagDIG seems to be experiencing a strong burst of star
formation which drives it to form stars at a rate 10 times larger than the
average of its entire life. This picture is frequently found in
galaxies classified as dIrr:  (NGC 6822: Gallart et al. 1996b,c; Pegasus:
Aparicio Gallart, \& Bertelli 1997a, Gallagher et al. 1998; Sextans A:
Dohm-Palmer et al. 1997, Van Dyk, Puche \& Wong 1998; Antlia:
Aparicio et al. 1997c; DDO 187: Aparicio et al. 1999). This might imply
that an important bias could exist in the classification of dwarfs as
bona-fide dIrrs towards objects experiencing strong star formation
bursts.

\section{Global integrated properties of SagDIG}

A summary of the basic parameters of SagDIG is given in Table 2. Except where
otherwise stated, the data listed are from this paper. The parameters listed
in Table 2 are: (1,2) - equatorial coordinates of the galaxy center; (3) -
the standard angular dimensions; (4) - the heliocentric radial velocity (data
for the four first lines are from NED); (5) - the radial velocity with
respect to the Local Group centroid (Karachentsev \& Makarov 1996); (6,7) -
interstellar reddening and extinction (Schlegel et al. 1998); (8,9) -
integrated colour and apparent magnitude of the galaxy; (10) - the central
surface brightness in $V$; (11,12) - the apparent reddening-corrected
$I$-magnitude of the TRGB, and the median reddening corrected colour of the
RGB measured at $M_{\rm I_0}=-3.5$; (13) - the mean metallicity estimated
from $(V-I)_{0,-3.5}$ using the calibration by Lee et al. (1993); (14-17) -
the distance modulus and linear distance from the Milky Way, from the Local
Group centroid, and from M 31; (18,19) - the total absolute magnitude and the
standard linear diameter of the galaxy; (20,21) - the hydrogen
mass-to-luminosity ratio and the total (virial) mass-to-luminosity ratio, for
which the total luminosity in $B$ obtained through $(B-V)_{\rm T}=0.85\times
(V-I)_{\rm T}-0.20$ and the hydrogen and total mass from Lo et al. (1993)
have been used.

As the presented data show, SagDIG is one of the faintest, smallest
irregular systems in the Local Group. According to its values of $M_{\rm
HI}/L_{\rm B}$ and $M_{\rm vt}/L_{\rm B}$, SagDIG seems to be an usual
gas-rich dIrr. Its low central brightness and low luminosity follow the
common relation between these parameters for dwarf galaxies (Caldwell et
al. 1998). But the derived low mean metallicity of SagDIG displaces it from 
 the main sequences [Fe/H] vs. $M_{\rm V}$ and [Fe/H] vs.
$\mu_{\rm v}(0)$ for dwarf galaxies (Lee 1995; Grebel \& Guhathakurta
1999). The unusually low metallicity of SagDIG needs, of course, an
independent confirmation. But if correct, studying the gas in SagDIG would
become an interesting task because it could be the lowest
metallicity dIrr known, with $Z$ smaller than that of I~Zw~18 even.

Being at more than 1 Mpc both from the Milky Way
and from M31, SagDIG is apparently situated outside these two
subsystems of the Local Group. Together with Sex A, Sex B, NGC 3109,
Antlia and UGCA 438, it forms a scattered envelope of the Local Group
populated by dIrr galaxies. In fact, SagDIG seems to be a
rather isolated dwarf galaxy whose autonomous evolution proceeds
without strong tidal influence from massive neighbours.

\section{Conclusions}

We have presented $V$ and $I$ CCD photometry of $\sim1550$ stars in the
Local Group dIrr galaxy SagDIG. The colour--magnitude diagram shows a well
defined red giant branch, and a sequence of blue stars. Based on
the $I$ magnitude of the TRGB we have derived the distance modulus of
SagDIG to be $(m-M)_0=25.13\pm 0.2$ for an adopted extinction of
$A_{I}=0.22$.  From the mean colour of the RGB $(V-I)_{0,-3.5}=
1.22$, a mean metallicity [Fe/H] $=-2.45\pm 0.25$ is obtained. With this
low Fe abundance, SagDIG could be the lowest metallicity star forming
dIrr known.

The absolute total magnitude, the total colour, and the central surface
brightness of SagDIG are derived to be $M_{ V}=-11.74$,
$(V-I)^{\rm T}_0=0.49$ and $\mu_{V}(0)=23.9$ mag arcsec$^{-2}$. The
radial star density distribution is fitted by an exponential law of
scale length $27\farcs1$, corresponding to 140 pc.

The star formation history of SagDIG has been estimated from the number
of blue and red stars in the colour--magnitude diagram and using a
synthetic CMD as reference. The star formation rate has been evaluated
for the intervals of time 0.2--15 Gyr, 0.05--0.2 Gyr and 0--0.05
Gyr, for which  has been obtained $\bar\psi_{0.2-15}=(1.3\pm
0.1)\times 10^{-4}$ $M_\odot$ yr$^{-1}$, $\bar\psi_{0.05-0.2}=(6.6\pm
0.8)\times 10^{-4}$ $M_\odot$ yr$^{-1}$, and $\bar\psi_{0-0.05}=(13\pm
5)\times 10^{-4}$ $M_\odot$ yr$^{-1}$, respectively (see a summary in
Table 1). The current strong enhancement of the observed star formation rate
 is a common feature of galaxies classified as dIrrs.

\acknowledgements

We thank N. A. Tikhonov for a fruitful collaboration. We thank the anonymous
referee for his/her usefull comments, that help improving the paper. This
work was partially supported by INTAS-RFBR grant 95-IN-RU-1390. AA is
financially supported by the IAC (grant P3/94) and by the DGES of the Kingdom
of Spain (grant PB97-1438-C02-01). NASA's Extragalactic Database (NED) has
been used in our work.

\newpage
\begin{table}
\caption{Star Formation Rates of SagDIG for different age intervals (in Gyr)}
\begin{tabular}{lc} \hline 
 & SFR\\
\hline
$\bar\psi_{0.2-15}$ \hfill \hspace{6mm} ($10^{-4}$ $M_\odot$ yr$^{-1}$) \hspace{6mm} & $1.3\pm 0.1$ \\
$\bar\psi_{0.05-0.2}$ \hfill \hspace{6mm} ($10^{-4}$ $M_\odot$ yr$^{-1}$) \hspace{6mm} & $6.6\pm 0.8$ \\
$\bar\psi_{0-0.05}$ \hfill \hspace{6mm} ($10^{-4}$ $M_\odot$ yr$^{-1}$) \hspace{6mm} & $13\pm 5$ \\
$\bar\psi_{0.2-15}/A$ \hfill \hspace{6mm} ($10^{-9}$ $M_\odot$ yr$^{-1}$ pc$^{-2}$) \hspace{6mm} & $0.28\pm 0.02$ \\
$\bar\psi_{0.05-0.2}/A$ \hfill \hspace{6mm} ($10^{-9}$ $M_\odot$ yr$^{-1}$ pc$^{-2}$) \hspace{6mm} & $1.44\pm 0.17$ \\
$\bar\psi_{0-0.05}/A$ \hfill \hspace{6mm} ($10^{-9}$ $M_\odot$ yr$^{-1}$ pc$^{-2}$) \hspace{6mm} & $2.8\pm 1.1$ \\
\hline
\end{tabular}
\end{table}

\newpage
\begin{table}
\caption{Observed and derived properties of SagDIG}
\begin{tabular}{lc} \hline 
\hspace{10mm} Parameter & SagDIG \\
\hline
(1) \hspace{8mm} R.A.(1950) & $\rm 19^h27^m05^s.4$ \\ 
(2) \hspace{8mm} Dec.(1950) & $-17^\circ 46' 59''$ \\
(3) \hspace{8mm} $a\times b$ ($'$) & $2.9\times 2.1$ \\
(4) \hspace{8mm} $V_{\rm h}$ (km s$^{-1}$) & $-77$ \\ 
(5) \hspace{8mm} $V_{\rm o}$ (km s$^{-1}$) & $+23$ \\ 
(6) \hspace{8mm} $E(B-V)$  & 0.12 \\ 
(7) \hspace{8mm} $A_{V}$, $A_{I}$  & 0.38, 0.22 \\
(8) \hspace{8mm} $(V-I)_{\rm T}$  & 0.65 \\ 
(9) \hspace{8mm} $V_{\rm T}$  & 13.77 \\
(10) \hspace{8mm} $\mu_{\rm V}(0)$ (mag$('')^{-2}$& $23.9\pm 0.1$ \\
(11) \hspace{8mm} $I_{\rm TRGB,0}$  & $21.13\pm 0.15$ \\
(12) \hspace{8mm} $(V-I)_{0,-3.5}$  & $1.17\pm0.05$ \\
(13) \hspace{8mm} [Fe/H] & $-2.45\pm 0.25$ \\
(14) \hspace{8mm} $(m-M)_0$  & $25.13\pm 0.2$ \\ 
(15) \hspace{8mm} $D_{\rm MW}$ (Mpc) & $1.06\pm 0.10$ \\ 
(16) \hspace{8mm} $D_{\rm LG}$ (Mpc) & 1.17 \\ 
(17) \hspace{8mm} $D_{\rm M 31}$ (Mpc) & 1.34 \\ 
(18) \hspace{8mm} $M^{\rm T}_{V}$  & $-11.74$ \\ 
(19) \hspace{8mm} $A$ (kpc) & 0.89 \\
(20) \hspace{8mm} $M_{\rm HI}/L_{\rm B}$ ($M_\odot/L_\odot$) & 1.3 \\
(21) \hspace{8mm} $M_{\rm vt}/L_{\rm B}$ ($M_\odot/L_\odot$) & 6.8 \\
\hline
\end{tabular}
\end{table}

\begin{figure}
\centerline{\psfig{figure=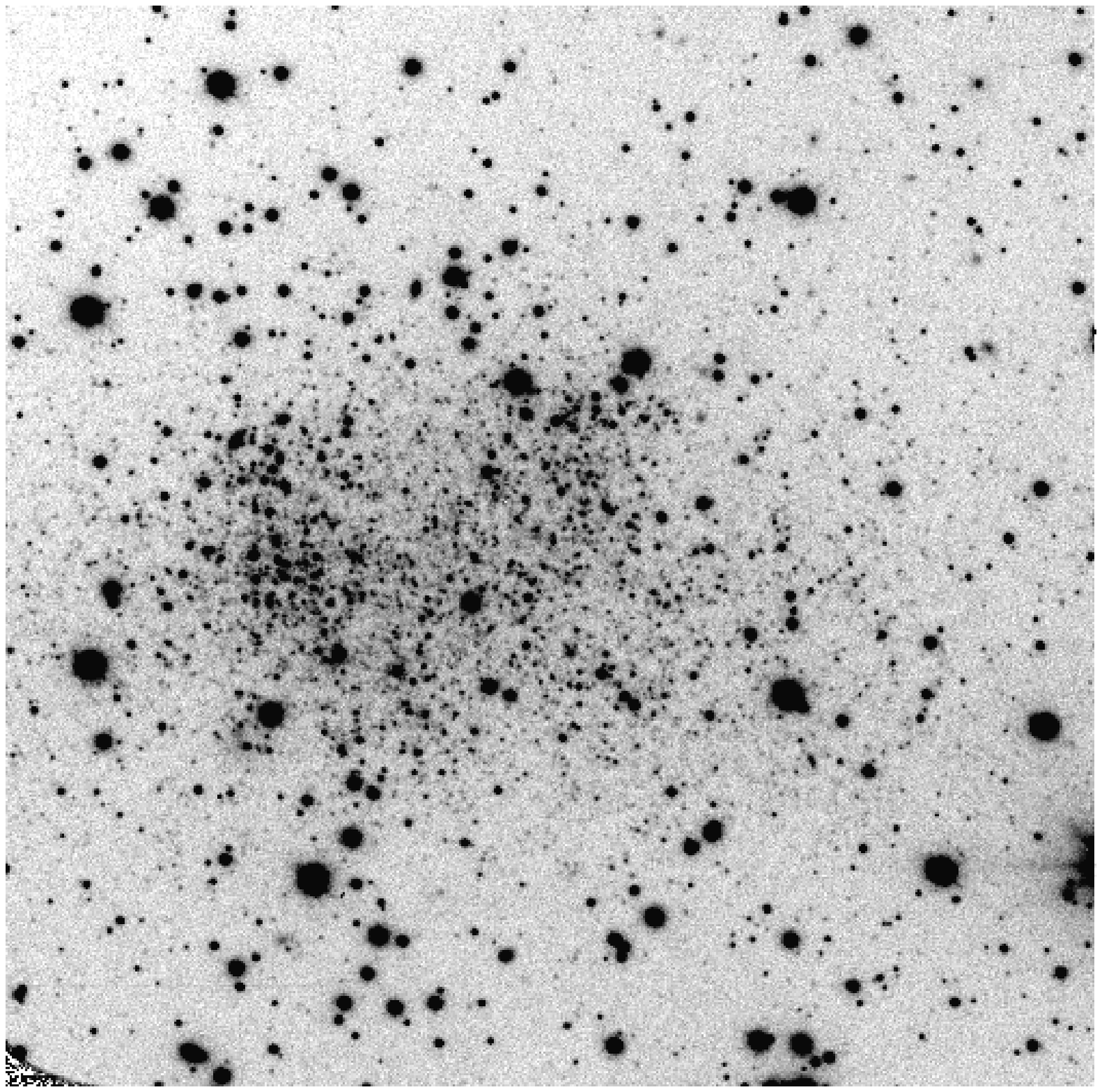}}
\caption{$V$ image of SagDIG obtained with the 2.56 m NOT with a
seeing of $\farcs 8$ (FWHM). Field size is $3\farcm 7\times 3\farcm 7$.
North is up, east is to the left.}
\end{figure}

\begin{figure}
\centerline{\psfig{figure=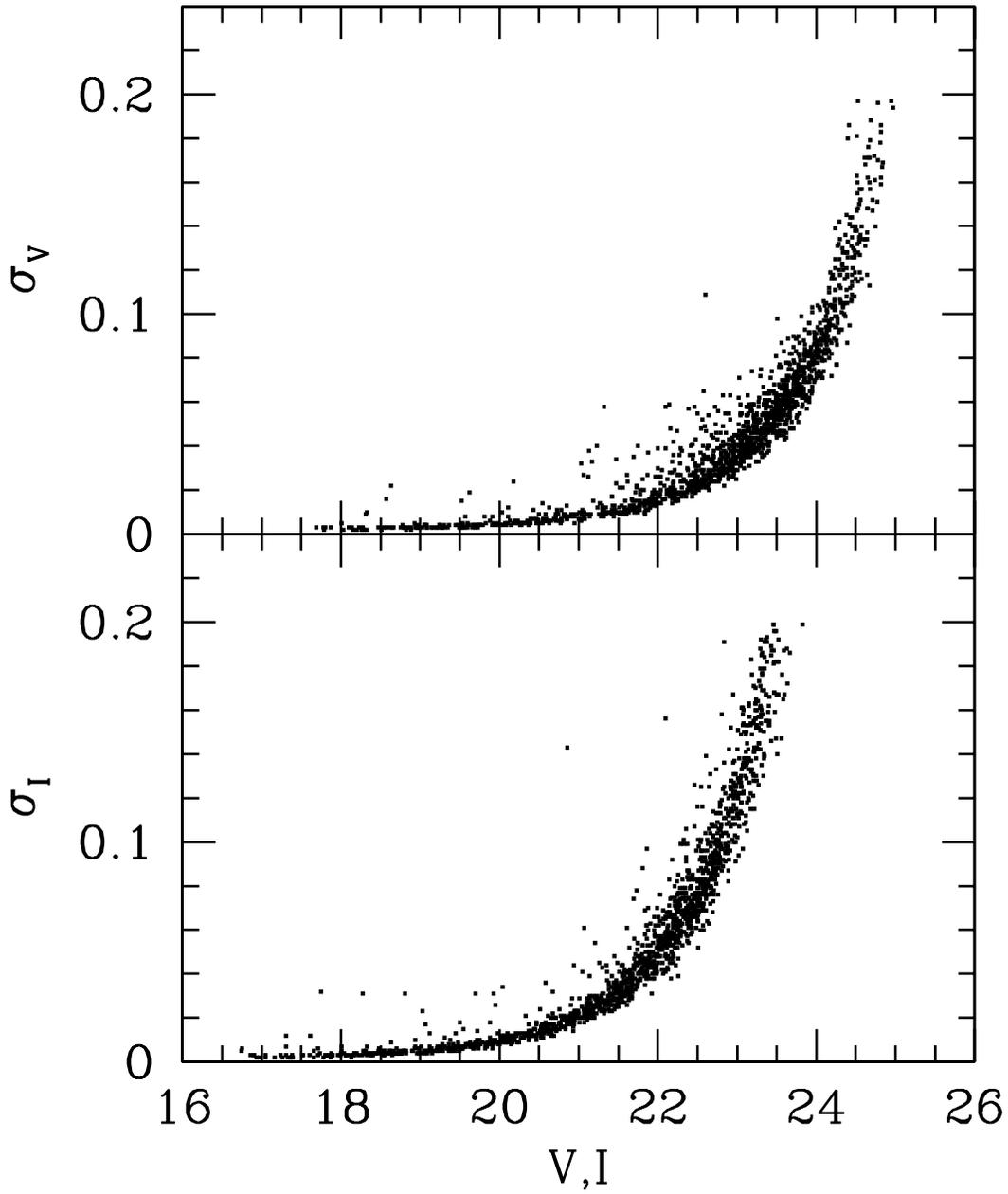}}
\caption{PSF fitting errors as given by ALLSTAR as a function of $V$ and $I$
magnitudes.}
\end{figure}

\begin{figure}
\centerline{\psfig{figure=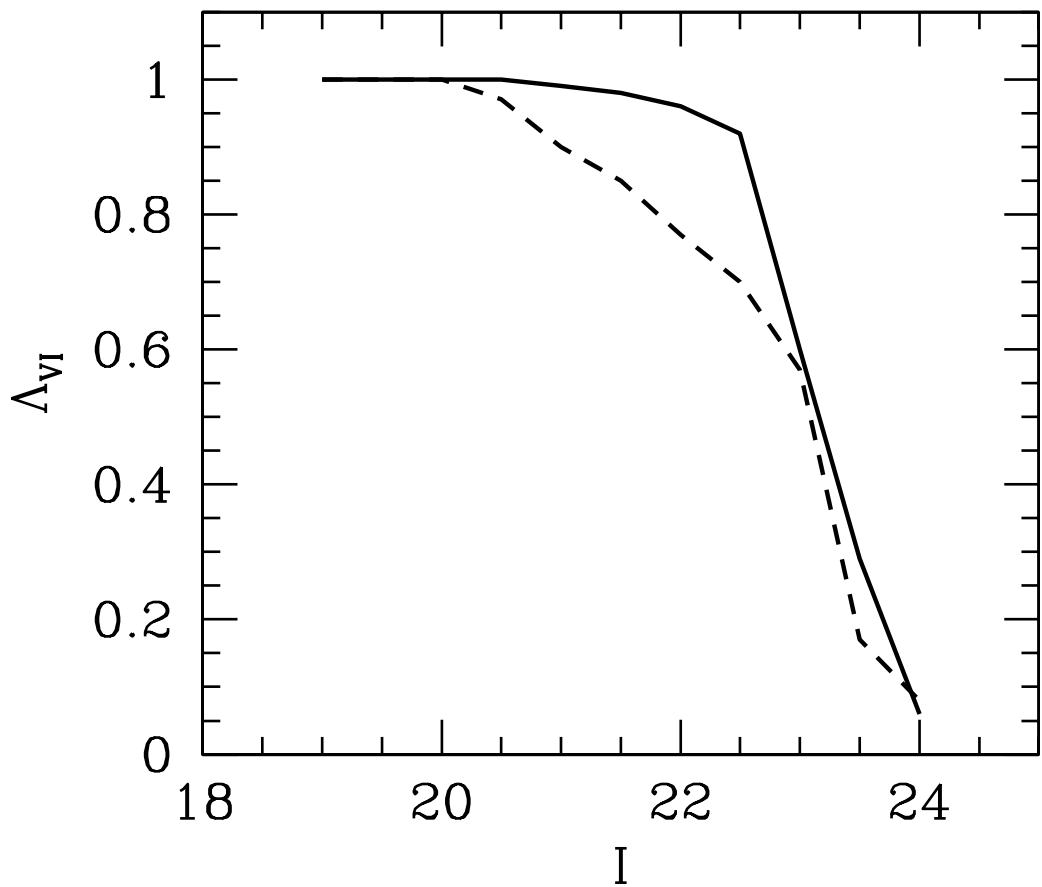}}
\caption{Completeness curves for blue [$(V-I)<0.6$; full line] and red
[$0.6<(V-I)<1.7$; dashed line] stars in the SagDIG field.}
\end{figure}

\begin{figure}
\centerline{\psfig{figure=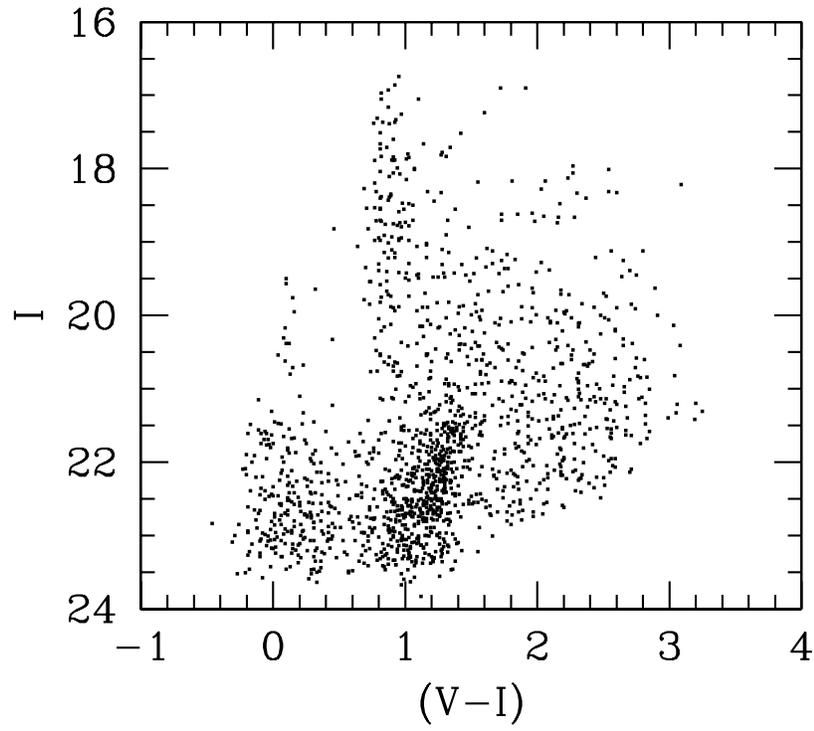,width=16cm}}
\caption{The colour--magnitude diagram of resolved stars in SagDIG field.}
\end{figure}

\begin{figure}
\centerline{\psfig{figure=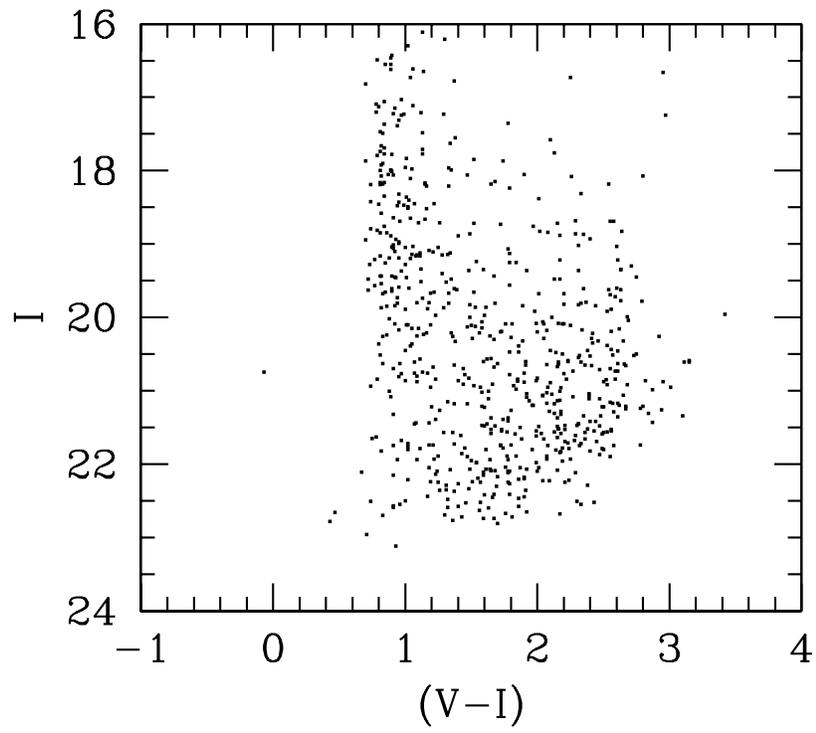,width=16cm}}
\caption{The colour--magnitude diagram of foreground stars.}
\end{figure}

\begin{figure}
\centerline{\psfig{figure=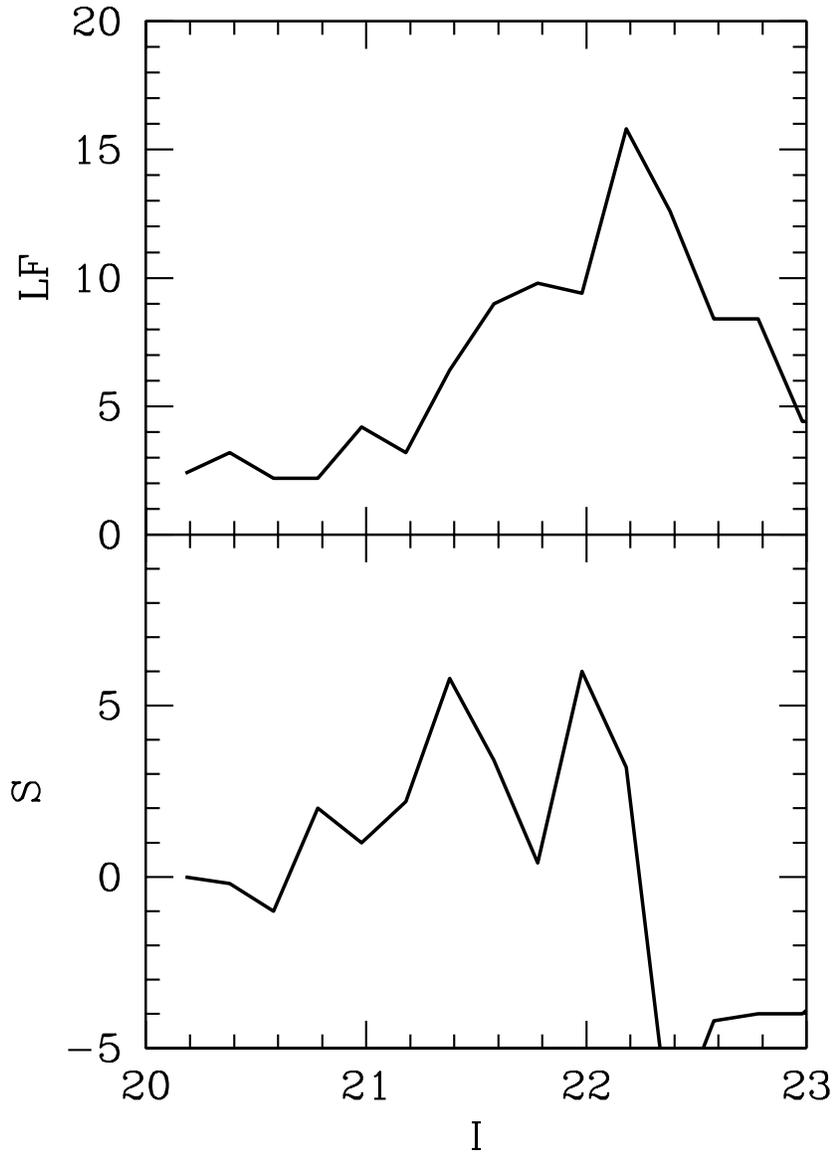,width=16cm}}
\caption{The luminosity function of red stars [$1.0<(V-I)<1.7$] (upper
pannel) and the same filtered through a Sobel filter of kernel [--1,0,+1]
(lower pannel). Stars in the central 1$'$ only have been used to reduce
foreground contamination effects.}
\end{figure}

\begin{figure}
\centerline{\psfig{figure=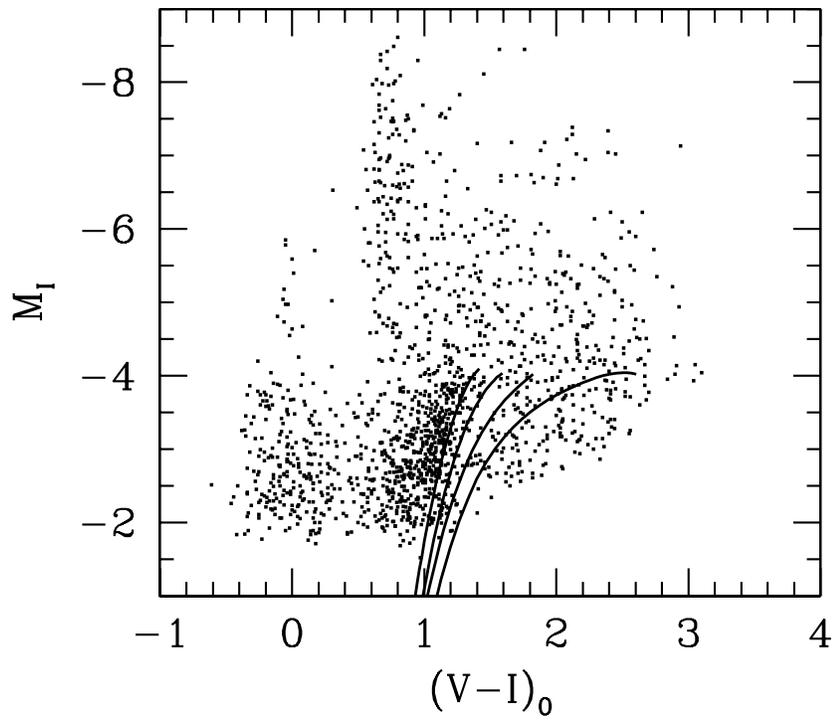,width=16cm}}
\caption{The globular cluster fidutial RGBs of Da Costa, \& Armandroff (1990)
overplotted to the reddening and distance-corrected CMD of SagDIG.}
\end{figure}

\begin{figure}
\centerline{\psfig{figure=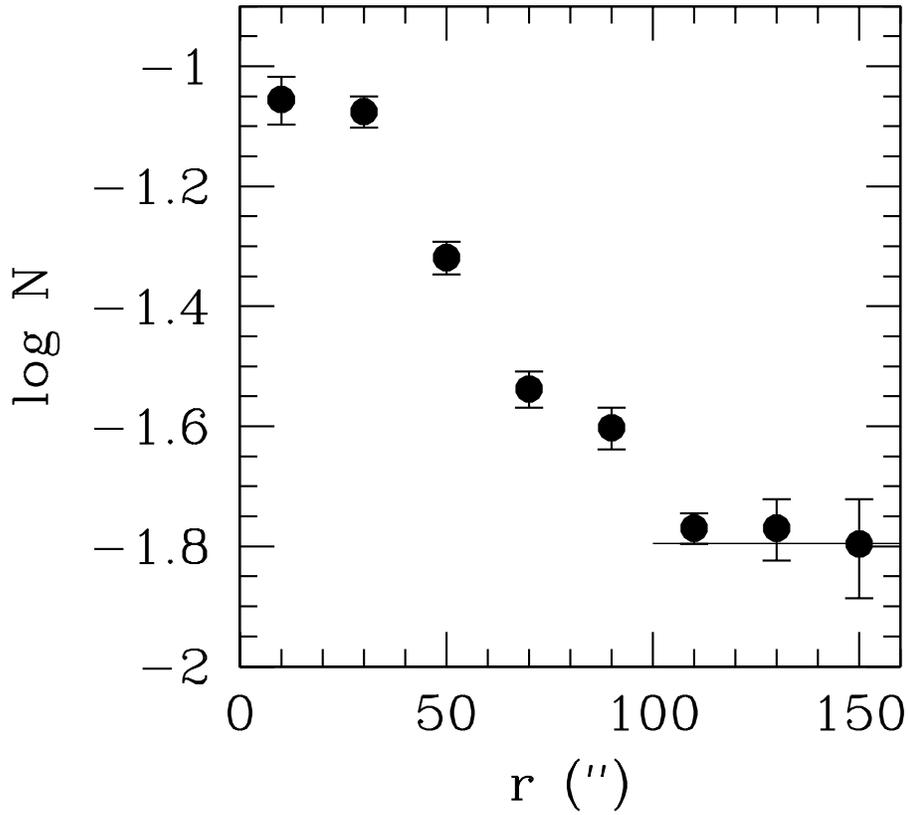,width=16cm}}
\caption{Radial density profile of SagDIG. Circular apertures have
been used. The adopted background level is marked by the horizontal
line in the  lower righthand corner.}
\end{figure}

\begin{figure}
\centerline{\psfig{figure=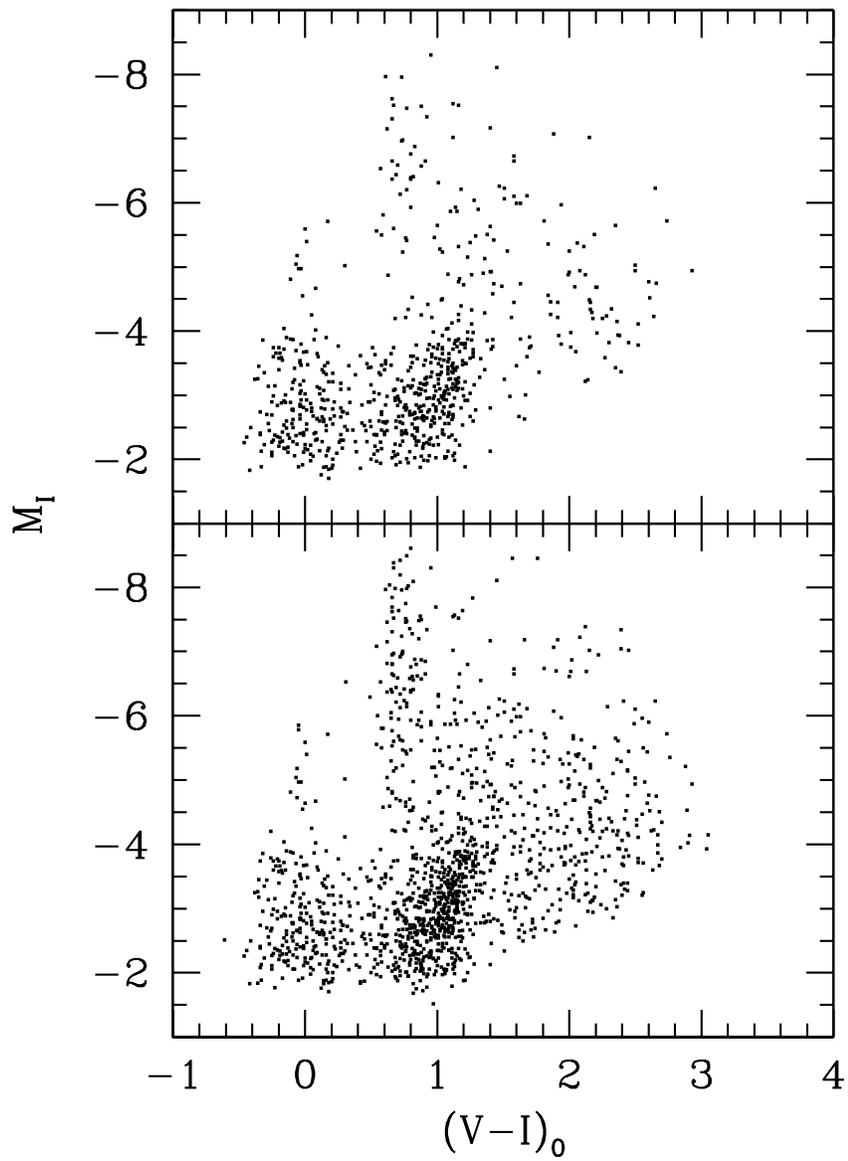,width=16cm}}
\caption{Reddening and distance-corrected CMDs for
the stars inside two circular regions centered in the galaxy center and radii
60$''$ and 120$''$.}
\end{figure}

\begin{figure}
\centerline{\psfig{figure=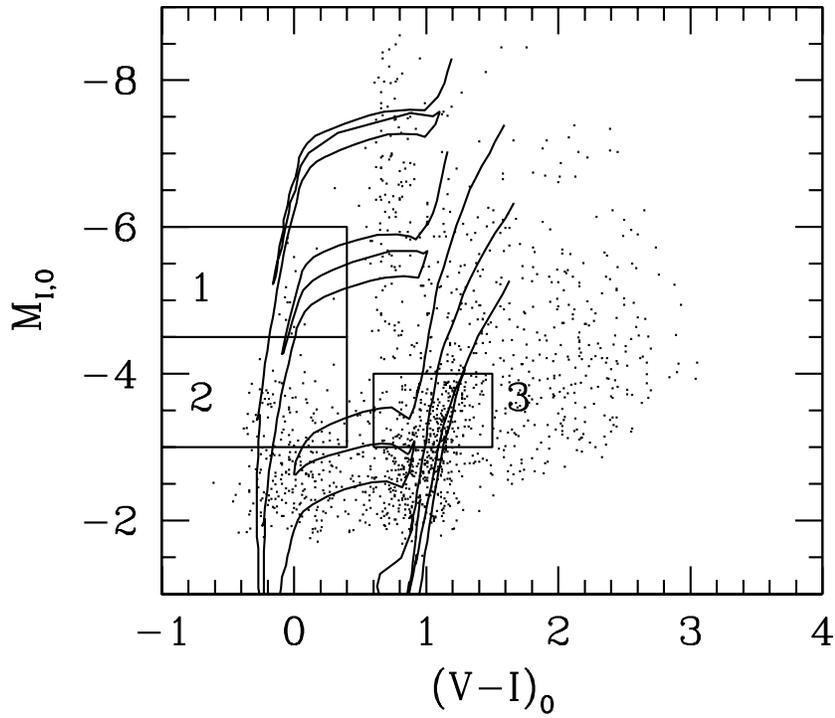,width=16cm}}
\caption{Reddening and distance-corrected colour--magnitude diagram for
the stars inside a circle of 2$'$ from the center of SagDIG. Five
isochrones from the Padua library with metallicity $Z=0.0004$ and ages
20 Myr, 50 Myr, 200 Myr, 1 Gyr and 10 Gyr are over-plotted together with
the three boxes used for the study of the SFH.}
\end{figure}

\begin{figure}
\centerline{\psfig{figure=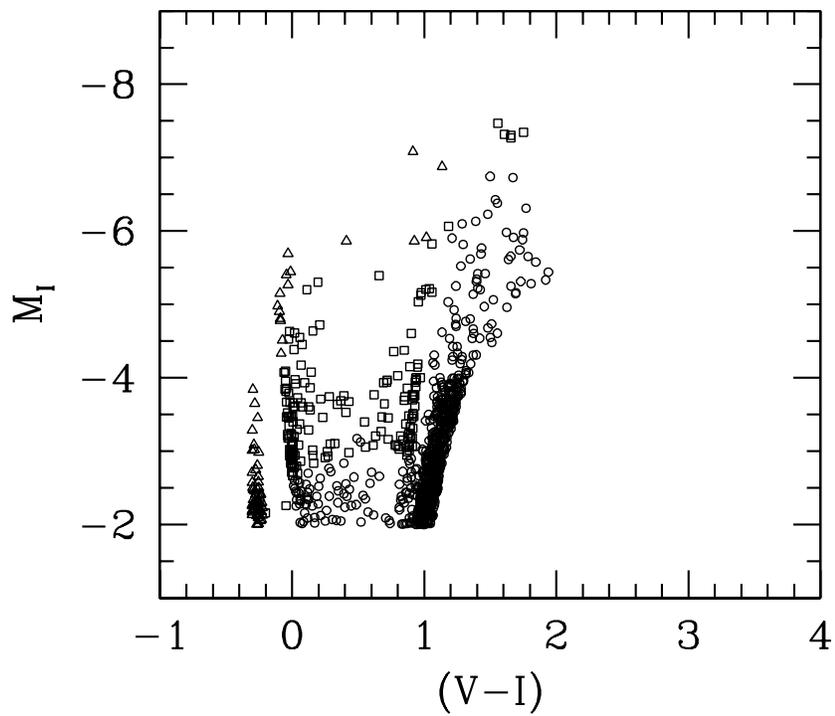,width=16cm}}
\caption{Synthetic CMD diagram computed using the SFH of SagDIG. Different
symbols correspond to different age intervals: triangles, 0--0.5 Gyr;
squares, 0.05--2 Gyr; circles, 0.2--15 Gyr.}
\end{figure}

\end{document}